\documentclass[fleqn,usenatbib,useAMS,twocolumn]{aastex63}
\usepackage{epstopdf}
\usepackage{epsfig,graphics,subfigure,psfrag,amsmath,amssymb,amscd}
\usepackage{url, hyperref, fancyhdr, multirow, aas_macros, makecell, verbatim} 

\usepackage{acronym}
\usepackage{CJK}
\usepackage{ulem, cancel} 

\def\apjl{Astrophys. J. Lett.}
\def\apjs{Astrophys. J. Suppl.}

\def\aap{Astron. Astrophys.}

\usepackage{lineno}

\let\oldequation\equation
\let\oldendequation\endequation
\renewenvironment{equation}{\linenomathNonumbers\oldequation}{\oldendequation\endlinenomath}

\let\oldalign\align
\let\oldendalign\endalign
\renewenvironment{align}{\linenomathNonumbers\oldalign}{\oldendalign\endlinenomath}

\let\oldgather\gather
\let\oldendgather\endgather
\renewenvironment{gather}{\linenomathNonumbers\oldgather}{\oldendgather\endlinenomath}

\newcommand{\KIAA}{Kavli Institute for Astronomy and Astrophysics, Peking University, 
Beijing 100871, China.}
\newcommand{\DOA}{Department of Astronomy, School of Physics, Peking University, Beijing 100871, China.
\href{Corresponding author.}{xian.chen@pku.edu.cn} }

\renewcommand\thetable{\Roman{table}}
\newcommand{\D}{\mathrm{d}}

\acrodef{GW}{gravitational-wave}
\acrodef{EM}{electromagnetic}
\acrodef{BH}{black hole}
\acrodef{BBH}{binary black hole}
\acrodef{BNS}{binary neutron star}
\acrodef{SNR}{signal-to-noise ratio}
\acrodef{AGN}{active galactic nucleus}
\acrodefplural{AGN}{active galactic nuclei}
\acrodef{CL}{confidence level}

\usepackage{orcid}

\begin{document}

\begin{CJK*}{UTF8}{gbsn} 



\title{Evidence of a fraction of LIGO/Virgo/KAGRA events coming from active galactic nuclei}

\author{Liang-Gui Zhu ({\CJKfamily{gbsn}朱良贵})
\orcidlink{0000-0001-7688-6504} 
}
\affiliation{\KIAA}

\author{Xian Chen ({\CJKfamily{gbsn}陈弦})
\orcidlink{0000-0003-3950-9317} 
}
\affiliation{\DOA}
\affiliation{\KIAA}

\date{\today}
%
\begin{abstract}

The formation channels of the gravitational-wave (GW) sources detected by
LIGO/Virgo/KAGRA (LVK) remain poorly constrained. Active galactic nucleus (AGN)
has been proposed as one of the potential hosts, but the fraction of GW events
originating from AGNs has not been quantified. Here, we constrain  the
AGN-origin fraction $f_{\rm agn}$ by analyzing the spatial correlation between
GW source localizations ($O1\!-\!O4$a) and AGNs (SDSS DR16).  We
report preliminary evidence of an excess of lower-luminosity ($10^{44.5} \lesssim L_{\rm bol} \le
10^{45}~\!\mathrm{erg~s}^{-1}$) as well as lower-Eddington ratio ($0.01 \lesssim \lambda_{\rm
Edd} \le 0.05$) AGNs around the LVK events, the explanation of which requires
$f_{\rm agn} = 0.39^{+0.41}_{-0.32}$ and $0.29^{+0.40}_{-0.25}$ (90\%
confidence level) of the LVK events originating from these respective AGN
populations.  Monte Carlo simulations confirm that this correlation is unlikely
to arise from random coincidence, further supported by anomalous variation of
the error of $f_{\rm agn}$ with GW event counts. These results support the
theoretical speculation that some LVK events come from lower-luminosity
or lower-accretion-rate AGNs, offering critical insights into the environmental
dependencies of the formation of GW sources. 

\end{abstract}


\section{Introduction}     \label{sec:introduction}

The network of ground-based \ac{GW} detectors, after four
observing runs, has observed more than three hundred events, most of which are
merging \acp{BBH} \citep{2019PhRvX...9c1040A, 
2021PhRvX..11b1053A, 2023PhRvX..13d1039A, 2024PhRvD.109b2001A, LVK_plans}. 
However, how these \acp{BBH} form is still under debate and far from
conclusive. The conventional idea is that astrophysical \acp{BBH} form in either
dense star clusters, or isolated binaries or stellar multiples 
\citep{1993MNRAS.260..675T, 2000ApJ...528L..17P, 
2020ApJ...900L..13A, 2023PhRvX..13a1048A}.
The detected \acp{BBH} could also be primordial \acp{BH} produced in the early universe
\citep[e.g.][]{sasaki16,ali17,inomata17,chen18,deluca20}.

An alternative idea is that BBHs could form in the accretion disks of \acp{AGN} \citep[e.g.,][]{cheng99,mckernan12,bartos17gas,ford22}.  In particular,
various types of hydrodynamical interactions can
help stellar-mass \acp{BH} accumulate at special locations in the disk
\citep{2014MNRAS.441..900M, 2016ApJ...819L..17B, 2021ApJ...911..124L, peng21, 2024MNRAS.530.2114G, 
2025ApJ...982L..13G}. 
Then the dense gaseous environment can assist the
pairing \citep{2023ApJ...944L..42L, 2023MNRAS.523.1126D} and hardening of the BBHs
\citep[e.g.][]{bartos17gas,stone17,yang19,lai23}. An interesting feature which may
differentiate the AGN channel from the rest is the hypothetical electromagnetic (EM)
flare that could be produced during the merger of BBHs in \acp{AGN}
\citep{2019ApJ...884L..50M,graham20flare,wang21ams,graham23}. But finding this EM counterpart is challenging
not only because AGNs are intrinsically 
highly luminous (which may mask the BBH merger's EM flare) and 
highly variable, but also because
the sky area and
redshift range inferred from each \ac{GW} event have relatively large errors.  
The latter
uncertainty results in a large number of AGNs within the ``error volume'',
sometimes as many as $10^5$ \citep[e.g.,][]{graham20flare, graham23, 2024PhRvD.110l3029C}. 


To overcome the limitation of localization accuracy, a statistical method
has been proposed \citep{bartos17method}. This method tests whether
there is an excess of some rare type of AGNs in the error volumes of the
detected BBHs. Its effectiveness has been verified by Monte Carlo
simulations \citep{veronesi22}.  Application of this method to real \ac{GW} data
already leads to meaningful results, which indicate that the most luminous AGNs
(e.g., $>10^{44.5}~{\rm erg~s^{-1}}$) cannot contribute more than $21\%$ of the
detected BBHs \citep{veronesi23obs,veronesi24obs}. 

Here we aim at strengthening the constraint by improving the method in two
aspects.  Observationally, AGNs are unevenly distributed in the sky and in a
wide range of redshift, but the previous works assumed a uniform number density
when calculating the likelihood of AGN excess.  This caveat can be addressed by
using a three-dimensional (3D) Voronoi tessellation method
\citep{1992stca.book.....O, 1994A&A...283..361V, Sochting:2001tp}, which has
been successfully applied in cosmological studies to investigate the
large-scale structure of the universe \citep{vandeWeygaert:2007ir,
2021inas.book...57V}.  Theoretically, luminosity is not the only factor that
determines the formation and merger rate of BBHs in AGNs.  Also important is
the accretion rate relative to the Eddington limit \citep[e.g.,][]{yang19},
since this ``Eddington ratio'' more directly determines the hydro and thermal
dynamical properties of an accretion disk \citep{2016ASSL..437.....K}.
Therefore, one should also select AGNs according to their Eddington ratios and
test their spatial correlations with BBHs. In the following, we show that the
improved method indeed reveals evidence that a fraction of BBHs are
coming from AGNs. 

The paper is organized as follows.  Section~\ref{sec:data} describes the data
used in this work, and  Section~\ref{sec:methodology} introduces our framework
of statistical analysis.  The results are presented in
Section~\ref{sec:results}, followed by a discussion in
Section~\ref{sec:discussion}.  Throughout this work, we adopt a flat $\Lambda$
cold dark matter ($\Lambda$CDM) cosmology with $H_0 =
67.9~\!\mathrm{km\,s^{-1}\,Mpc^{-1}}$ and $\Omega_M = 0.3065$
\citep{2016A&A...594A..13P}, consistent with the cosmological parameters used
in the GW event catalogs \citep{2019PhRvX...9c1040A, 2021PhRvX..11b1053A,
2023PhRvX..13d1039A, 2024PhRvD.109b2001A}.

\section{GW and AGN Data}   \label{sec:data}

The data used in our analysis consist of skymaps of \ac{GW} sources and
AGN catalogs.  The skymaps of \ac{GW} sources are obtained from the
latest data published by the LIGO/Virgo/KAGRA (LVK) network, including two
datasets: (1) the officially published GW transient catalogs from the first
three observing runs ($O1$ to $O3$) \citep{2019PhRvX...9c1040A,
2021PhRvX..11b1053A, 2023PhRvX..13d1039A, 2024PhRvD.109b2001A} \footnote{Data
of LVK's $O1-O3$ GW candidates are available at:
\url{https://gwosc.org/eventapi/html/GWTC/}. }, and (2) the preliminarily
released GW alerts with significant detections from the fourth observing run
($O4$) 
\footnote{The released skymaps of $O4$ GW alerts are available at:
\url{https://gracedb.ligo.org/superevents/public/O4/}. }, as of January 19,
2025. These skymaps provide us with the probability distribution function 
$p(\mathbf{x})$ of the 3D localization of each GW event. 

The AGN catalog used in this work is from the sixteenth data release (DR16) of
the Sloan Digital Sky Survey (SDSS) \citep{2020ApJS..250....8L}, which contains
approximately $3 \times 10^5$ AGNs with $z<1.5$.  We adopt the bolometric
luminosities ($L_{\rm bol}$, in units of $\rm{erg~s^{-1}}$ throughout) 
and Eddington ratios ($\lambda_{\rm Edd}$)
published in \cite{2022ApJS..263...42W} \footnote{The re-published AGN catalog
of SDSS DR16 are available at:
\url{http://quasar.astro.illinois.edu/paper_data/DR16Q/}. }, which have taken into
account bolometric corrections \citep{Richards:2006xe} and used three different
recipes to calibrate the massive 
\ac{BH} masses $M_{\rm MBH}$ \citep{Shen:2010aa}.  To
ensure a relatively complete AGN catalog, we exclude AGNs from the regions with a sky surface 
density below $2~{\rm deg}^{-2}$.  The final \ac{AGN} catalog covers
approximately $26\%$ of the entire sky. Additional details can be found in
Appendix \ref{sec:appendix_data}. 

To exclude the GW events that are poorly localized or residing in the sky regions
where AGN catalogs are highly incomplete, we impose three more criteria. 
(i) The comoving volume $\Delta
V_{\rm c}$, within which a GW source is localized with a $90\%$ \ac{CL}, is
smaller than $10^{11}~\!{\rm Mpc}^3$. (ii) Over $20\%$ of this ``error volume''
($\Delta V_{\rm c}$) is covered by the SDSS AGN survey.  (iii) The entire error
volume is confined to redshifts below $z=1.5$.  After applying these criteria
to a total of $93$ events reported by $O1-O3$, we find $29$ GW events
satisfying our criteria.  For the ongoing $O4$ run, we consider the first $191$
significant GW candidates with an astrophysical origin probability of $p_{\rm
astro} > 0.9$, of which $65$ satisfy our selection criteria.  We note that the
skymaps of the $O4$ GW candidates have been provided by \textsf{Bayestar}
\citep{Singer:2015ema} or \textsf{Bilby} \citep{Ashton:2018jfp} pipelines.  We
use the \textsf{Bayestar} skymaps for all \ac{GW} candidates, since only a
fraction of \ac{GW} candidates have \textsf{Bilby} skymaps.  Further details on
the \ac{GW} datasets are provided in Appendix \ref{sec:appendix_data}. 

\begin{figure}[t]
 \centering
 \includegraphics[width=0.47\textwidth]{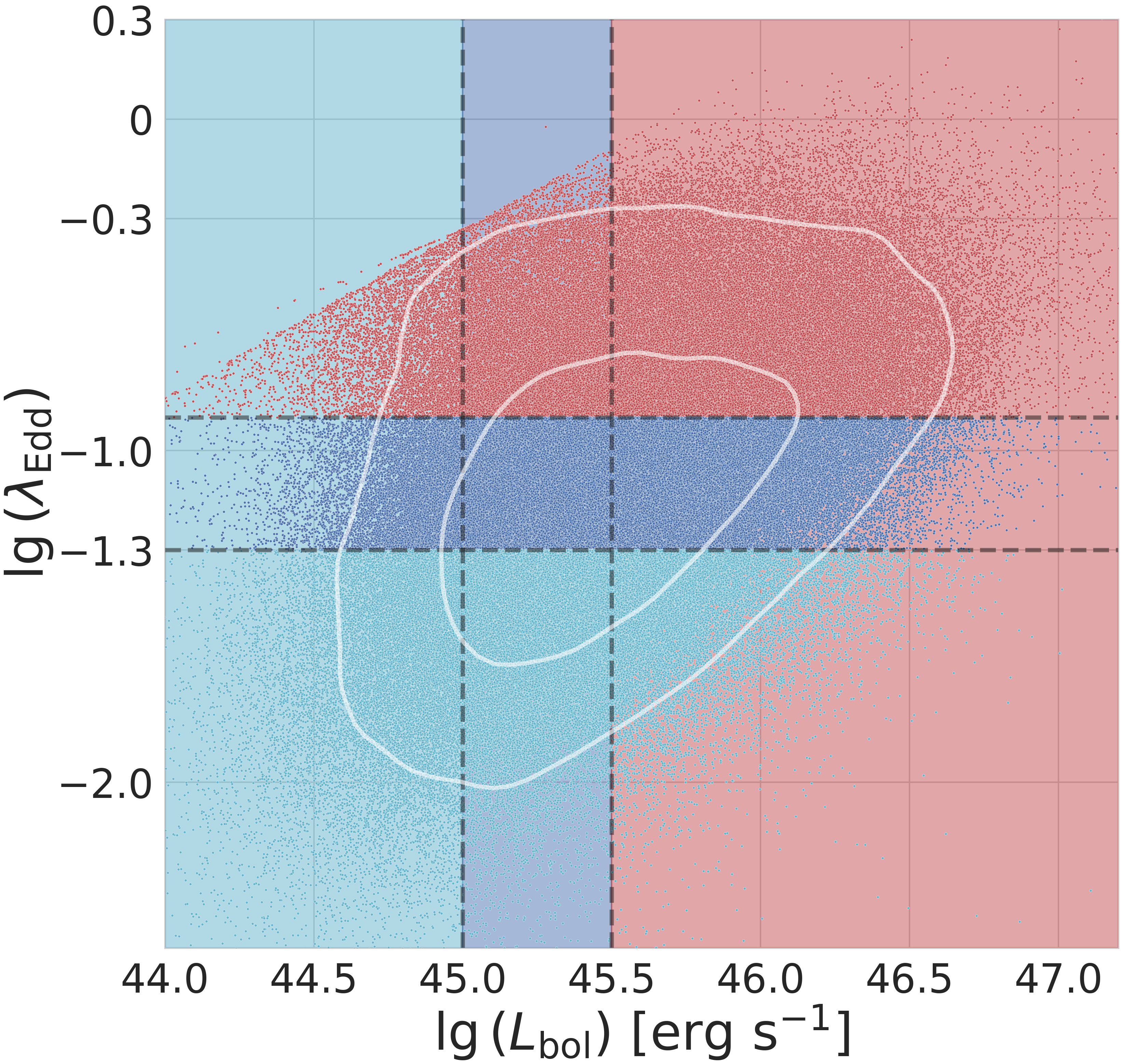}
 \caption{Distributions of AGNs in the $L_{\rm bol} - \lambda_{\rm Edd}$ plane.
 The two white contours represent the 50\% and 90\% CLs of the AGN distribution. 
 The cyan, blue, and red regions (dots) correspond to the lower, moderate and higher $L_{\rm bol}$ ($\lambda_{\rm Edd}$) AGN sub-catalogs, respectively.   }
 \label{fig:Lbol_EddR}
\end{figure}

To investigate the spatial correlations between GW events and \acp{AGN} of
different bolometric luminosities $L_{\rm bol}$ or Eddington ratios
$\lambda_{\rm Edd}$,  we divide the full AGN catalog into three sub-catalogs
according to either $L_{\rm bol}$ or $\lambda_{\rm Edd}$. 
The sub-catalogs are as follows:
\begin{itemize}
	\item{According to bolometric luminosity: lower ($\lg L_{\rm bol}  \leq 45$), 
       moderate ($45 < \lg L_{\rm bol} \leq 45.5$), and 
       higher ($\lg L_{\rm bol} > 45.5$) $L_{\rm bol}$ sub-catalogs; } 
 \item{According to Eddington ratio: lower ($\lg \lambda_{\rm Edd} \leq -1.3$), 
       moderate ($-1.3 < \lg \lambda_{\rm Edd} \leq -0.9$) and 
       higher ($\lg \lambda_{\rm Edd} > -0.9$) $\lambda_{\rm Edd}$ sub-catalogs.}
\end{itemize}
The division is partly based on a comparable number of AGNs in different
sub-catalogs and partly motivated by the predictions from previous theoretical
works \citep{bartos17gas, stone17, yang19, yang20, peng21, Delfavero:2024clh, 
2024MNRAS.530.2114G, 2025ApJ...982L..13G}.
Figure \ref{fig:Lbol_EddR} shows the resulting sub-catalogs 
\footnote{All data used in this work are available at:
\url{https://zenodo.org/records/15387462}. }.  
Notice that the sub-catalogs classified by bolometric luminosity may overlap with those
classified by Eddington ratio. For example, about $40\%$ of the AGNs in the
lower-$L_{\rm bol}$ sub-catalog also fall in the lower-$\lambda_{\rm Edd}$ sub-catalog.
Additionally, we further highlight that due to selection effects, 
the AGN samples in the lower-$L_{\rm bol}$ and lower-$\lambda_{\rm Edd}$ sub-catalogs predominantly occupy 
the bolometric luminosity range of $44.5 \lesssim \lg L_{\rm bol}  \leq 45$ and 
the Eddington ratio range of $-2 \lesssim \lg \lambda_{\rm Edd} \leq -1.3$, respectively.

\section{Methodology}  \label{sec:methodology}

We adopt the method proposed by \citet{bartos17method} to statistically test
the spatial correlation between the LVK events and the SDSS AGNs. This method combines the localization information of a GW
source and the AGNs inside the error volume to compute the values of two
probability distribution functions $\mathcal{S}$ and $\mathcal{B}$, known as
the ``signal probability'' and ``background probability'', respectively. These
functions are constructed in a way such that the statistical expectation of
$\mathcal{S}$ is greater than that of $\mathcal{B}$ if the GW source comes from
an AGN, and vice versa. 

In the real situation where a number of $N$ GW events
(mainly BBHs) are detected and only a fraction of them are from AGNs, one should
compute $\mathcal{S}_i$ and $\mathcal{B}_i$ for each GW source and evaluate the
total likelihood
\begin{align}  \label{eq:likeli_tot}
\!\!\!\!\!\!\!\!\!\!\!\!\!\!
\mathcal{L}(f_{\rm agn}) \!=\!  \prod_{i=1}^{N} \! \bigg[ 0.9 \cdot c_i \cdot\! f_{\rm agn} \cdot \mathcal{S}_i
 + \big( 1 - 0.9 \cdot c_i \cdot\! f_{\rm agn} \big) \!\cdot \mathcal{B}_i    \bigg].
\end{align}
Here, $f_{\rm agn}$ is the hypothesized fraction of BBHs from AGNs,
$c_i$ accounts for the completeness of the AGN catalog since faint objects
may be missed by observations, and the coefficient $0.9$ comes from
the credibility of GW event's sky localization. 
Previous mock simulations conclude that the peak of this likelihood function 
agrees with the true value of $f_{\rm agn}$ \citep{veronesi23obs, zhu24emri}.
Later in Appendix \ref{sec:appendix_method} we will prove that
the previous conclusion is valid when the number of GW events is large
and the random fluctuation of local AGN number density is small.

In the original proposal of \citet{bartos17method}, AGNs are assumed to
follow a Poisson distribution in the sky \citep[also in the mock simulation
by][]{veronesi22}. Later work relaxes this assumption, and uses the real sky
positions of AGNs as well as the probability density of 3D localization [the
aforementioned $p_i(\mathbf{x})$, with the subscript denoting the $i$th LVK
event] to construct $\mathcal{S}_i$ and $\mathcal{B}_i$ \citep{veronesi23obs}.
This improved method, however, requires that (i) the spatial density of AGNs
($n_{\rm agn}$) is constant and (ii) the error volumes of GW events are fully
covered by AGN surveys, which do not agree with real data. 

\begin{figure*}[t]
 \centering
 \includegraphics[width=0.99\textwidth]{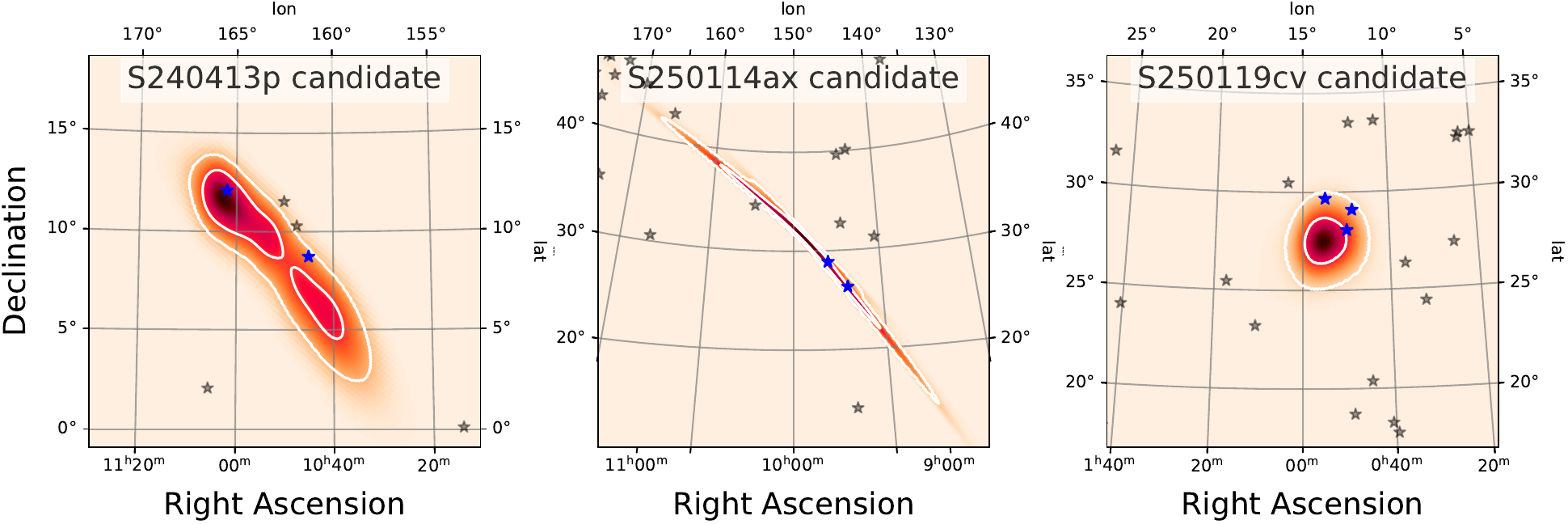}
 \caption{Skymaps of three precisely localized $O4$ GW candidates and scatter plots of their neighboring AGNs. 
 The two white contours in each panel represent the 50\% and 90\% CLs of the skymap. 
 The blue/gray stars represent the AGNs inside/outside the localization error volume of 
90\% CL.   }
 \label{fig:gw_skymaps}
\end{figure*}

To allow $n_{\rm agn}$ to vary, we rewrite the signal probability as
\begin{equation}  \label{eq:prob_Si}
\mathcal{S}_i = \sum_{j=1}^{N_{{\rm agn}}} \frac{ p_i(\mathbf{x}_j) }{n_{\rm agn}(\mathbf{x}_j) }, 
\end{equation}
where $\mathbf{x}_j$ denotes the 3D position of an AGN which appears in the
error volume of the $i$th LVK event, and $N_{{\rm agn}}$ is the total
number of the cataloged AGNs which are present in the 90\% CL 
error volume.  
Figure~\ref{fig:gw_skymaps} illustrates the sky localization probability distributions 
and the surrounding AGN distributions for three precisely localized GW events. 
The blue stars mark the AGNs falling
 within the $90\%$ confidence contours and hence are
included 
in the summation of Equation (\ref{eq:prob_Si}). 
To derive $n_{\rm agn}$ as a function of $\mathbf{x}_j$, we apply a 3D first-order
\emph{Voronoi tessellation} method \citep{voronoi1908nouvelles, 1992stca.book.....O,
brakke2005statistics, 2009JSP...134..185L}. 
The Voronoi tessellation method has found broad application in cosmological studies of 
large-scale structure in galaxy spatial distribution 
\citep{1994A&A...283..361V, Sochting:2001tp, vandeWeygaert:2007ir, 2021inas.book...57V}.
In our analysis, we utilize this method to partition the comoving volume covered by 
the SDSS AGN catalog into polyhedral cells 
within a Cartesian coordinate system, with each cell containing exactly one AGN. 
We also use \textsf{alphashape} to determine sky-coverage boundaries 
based on the spatial distribution of the full AGN catalog. Isolated AGNs and 
those in small-scale clustering regions outside the boundaries 
were explicitly excluded, to minimize the error in calculating Voronoi volume. 
Then we use the volume of the Voronoi cell, $V_{\rm cell}$, to calculate
the \ac{AGN} density as $n_{\rm agn} = 1/V_{\rm cell}$. 

Equation~(\ref{eq:prob_Si}) no longer requires the error volumes to be fully
covered by AGN surveys because the equation only sums up those cataloged AGNs. 
But such partial coverage does result in a loss of the signal, so
we account for it by revising the completeness factor in the total likelihood as
\begin{equation}  \label{eq:compl}
c_i = \frac{1}{\mathcal{N}_{\rm CL}} \!\!\!\! \iiint\limits_{~~\Delta V_{\rm c, agn}} \!\!\!\! P_{\rm c}(\alpha, \delta, D_{\rm c}) p_i(\alpha, \delta, D_{\rm c}) J \D\alpha \D\delta \D D_{\rm c} ,
\end{equation}
where $(\alpha, \delta)$ are celestial coordinates, $D_{\rm c}$ is comoving distance, 
$P_c \le 1$ is a selection function depending on the sensitivities of observations as well as
the luminosities and clustering properties of AGNs
\citep[e.g.,][]{eBOSS:2017cqx, eBOSS:2020gbb, eBOSS:2020uxp}, 
$J \equiv D_{\rm c}^2 \cos\delta$ is Jacobian determinant 
for spherical-to-cartesian coordinate transformation, 
and $\mathcal{N}_{\rm CL} = 0.9$ normalizes the 90\% localization CL.  Notice that
the integration is perform not in the entire error volume, but in the portion
occupied by the cataloged AGNs, whose total volume is denoted as $\Delta V_{\rm
c, agn}$.

After the above modifications, Equation~(\ref{eq:prob_Si}) can be understood as
follows. If a GW event comes from an AGN, this host AGN should be found close
to the peak of $p_i(\mathbf{x})$.  Including it in the summation naturally
increases the statistical prominence of the signal $\mathcal{S}_i$.  In
addition, the summation is weighed by $1/n_{\rm agn}$, the reciprocal of AGN
density, to lower the expectation of finding real signals where AGNs are so
abundant that a random coincidence with GW events is probable.

We also modify the expression of the background probability according to the
previous considerations.
By definition, $\mathcal{B}_i$ is the statistical expectation of 
Equation~(\ref{eq:prob_Si}) when there is no spatial correlation between GW events
and AGNs, and its value is $0.9$ in the previous works \cite[e.g.][]{veronesi23obs,veronesi24obs}. In our case, a fraction of the error volume lies outside the region covered by our SDSS 
AGNs, so the expectation of Equation~(\ref{eq:prob_Si}) is smaller than $0.9$.
Therefore, we define a ``covering factor'',
\begin{align}  \label{eq:f_cover}
	f_{{\rm cover},i} = \frac{1}{\mathcal{N}_{\rm CL}} \iiint\limits_{\Delta V_{\rm c, agn}} p_i(\alpha, \delta, D_{\rm c}) J\D\alpha \D\delta \D D_{\rm c}, 
\end{align}
and the background probability is reduced to
\begin{align}  \label{eq:prob_Bi}
\mathcal{B}_i = 0.9 \cdot f_{{\rm cover},i}.
\end{align}

Those familiar with the earlier works \citep{veronesi23obs, veronesi24obs} may
notice that we have omitted a normalization factor of $1/\Delta V_{\rm c}$, the
reciprocal of the error volume, in front of both $\mathcal{S}_i$ and
$\mathcal{B}_i$. We can do this because the total likelihood depends on a
multiplication of these factors according to Equation~(\ref{eq:likeli_tot}),
which leads to a constant coefficient and does not affect the probability
distribution of $f_{\rm agn}$.

We notice that the two parameters $f_{\rm agn}$ and $c_i$ always appear as a
product in the likelihood. Therefore, they are degenerate in our analysis.
Breaking this degeneracy requires an accurate estimation of $c_i$, but it is
difficult due to the poor observational constraints on the luminosity function
\citep{2007ApJ...654..731H, 2020MNRAS.495.3252S, 2019MNRAS.488.1035K} and 
the Eddington-ratio distribution function \citep{2010A&A...516A..87S, 2022ApJS..261....9A} of AGNs,
especially about how these functions vary with redshift 
(we discuss the completeness of 
GW events under different AGN catalogs in Appendix \ref{sec:appendix_data}, 
the estimates are rough and were not used for the subsequent analysis in this work). 
Because of the uncertainty in $c_i$, in the following calculation we will assume 
$c_i = f_{{\rm cover},i}$.  Effectively, this means the selection function $P_{\rm
c}(\alpha, \delta, D_L)$ in Equation~(\ref{eq:compl}) is $1$. This is obviously
an optimistic assumption and will lead to an overestimation of $c_i$.
Consequently, the resulting $f_{\rm agn}$ is lower than its real value and
should be regarded as a conservative estimation.

\section{Results}   \label{sec:results}

\subsection{Evidence of a non-zero $f_{\rm agn}$}   \label{sec:result_preciLoc}

\begin{figure}[t]
 \centering
 \includegraphics[width=0.47\textwidth]{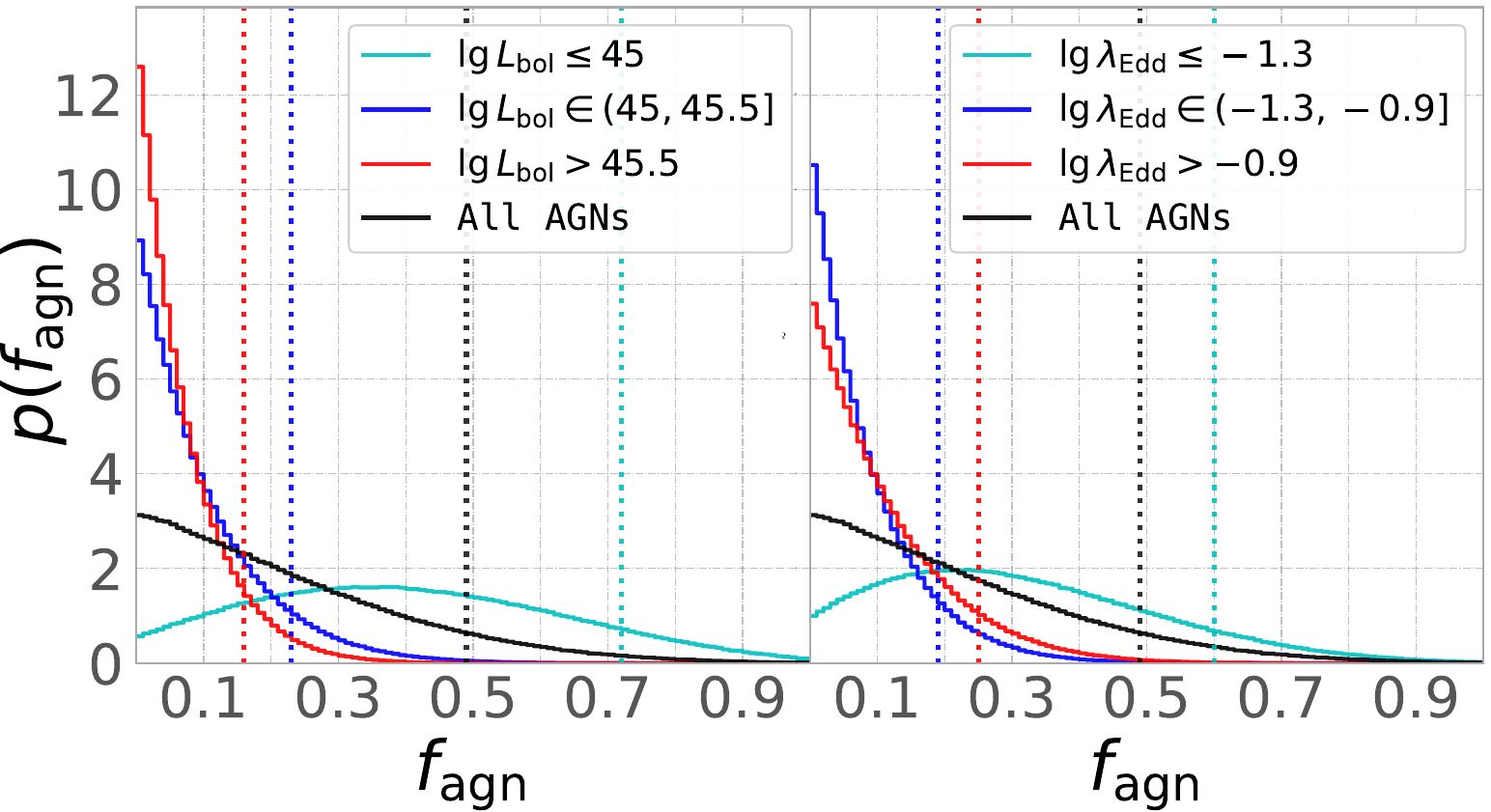}
 \caption{Probability distributions of $f_{\rm agn}$ derived for 
  all $O1\!-\! O4$a GW events.
 Different curves correspond to lower (cyan), moderate (blue),  and higher (red) $L_{\rm bol}$ (left panel) 
 or $\lambda_{\rm Edd}$ (right) sub-catalogs. 
Results for the full catalog are shown with black curves. 
 Vertical dotted lines mark the 90\% CL upper limits.  }
 \label{fig:fagn_PDF_allLocGW}
\end{figure}

We first use all GW events from the LVK network's $O1\!-\!O3$ and $O4$ runs to
constrain $f_{\rm agn}$.  The outcome depends on the choice of sub-catalog of
AGNs defined in Section~\ref{sec:data}.  Figure~\ref{fig:fagn_PDF_allLocGW}
shows the results. If we use moderate/higher-$L_{\rm bol}$ and
moderate/higher-$\lambda_{\rm Edd}$ sub-catalogs, the maximum likelihoods are
found at $f_{\rm agn}^{\rm best} = 0$ (see blue and red curves).  The same
result is found when we use the full AGN catalog (black curves).  Therefore,
correlation between GW events and AGNs is not favored in these four cases. 

However, when using the lower-$L_{\rm bol}$ 
($44.5 \lesssim \lg L_{\rm bol}  \leq 45$) 
and lower-$\lambda_{\rm Edd}$ 
($-2 \lesssim \lg \lambda_{\rm Edd} \leq -1.3$) 
sub-catalogs (cyan curves), the probability distributions of $f_{\rm agn}$
become significantly different and the most probable values are no longer
zero.  The median values with symmetric $90\%$ CLs become $f_{\rm agn} =
0.39_{-0.32}^{+0.41}$ and $f_{\rm agn} = 0.29_{-0.25}^{+0.40}$, respectively.
We notice that this is the first time that a non-zero most probable value of
$f_{\rm agn}$ is found by analyzing the spatial correlation between GW events
and AGNs. 

\begin{figure}[t]
 \centering
 \includegraphics[width=0.47\textwidth]{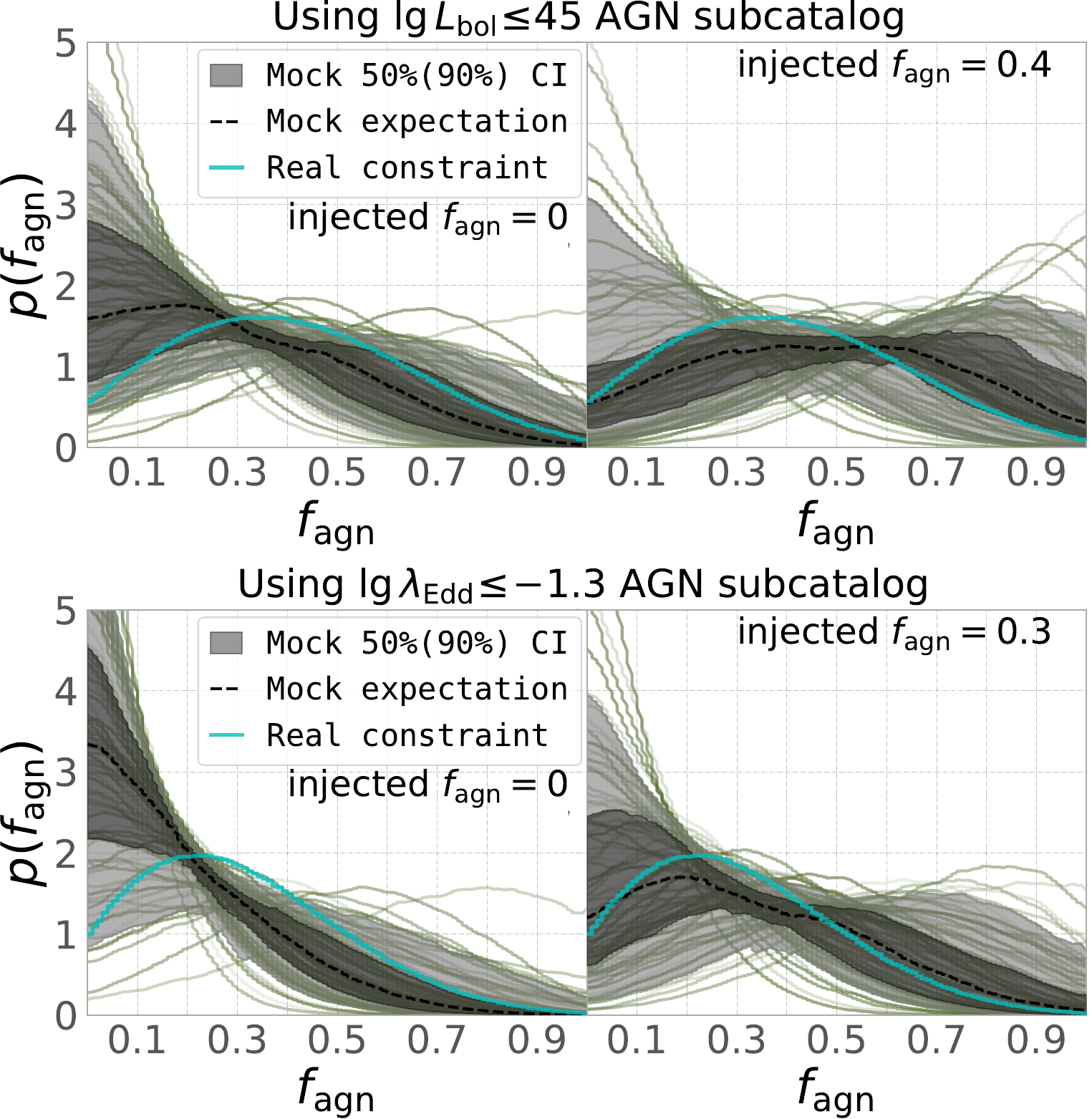}
 \caption{Comparison between real and mock constraints on $f_{\rm agn}$ 
 based on all $O1\!-\!O4$a GW events. 
 \textit{Top panels}: constraints using the lower $L_{\rm bol}$ sub-catalog with 
 injected values $f_{\rm agn}^{\rm inj} = 0$ and $f_{\rm agn}^{\rm inj} = 0.4$,
respectively. 
 \textit{Bottom panels}: constraints using the lower $\lambda_{\rm Edd}$ sub-catalog with 
 injected values $f_{\rm agn}^{\rm inj} = 0$ and $f_{\rm agn}^{\rm inj} = 0.3$. 
 In each panel, the cyan solid line correspond to the result derived from real 
observation, 
 while the gray curves correspond to mock simulation. 
The black dashed curve denotes the median of the simulated results, 
and the gray shaded areas indicate the 50\% and 
 90\% distribution intervals.   }
 \label{fig:fagn_PDF_mock}
\end{figure}

Before making any conclusion, we have to test whether the non-zero value of
$f_{\rm agn}$ could arise from statistical fluctuation. Such a test is
motivated by the following observations. (i) The $90\%$ CL corresponds to a
wide range of $f_{\rm agn}$ (see cyan curves in
Figure~\ref{fig:fagn_PDF_allLocGW}). (ii) The probability density at $f_{\rm
agn}=0$ is not zero.  (iii) The signal probability $\mathcal{S}_i$ according to
Equation~(\ref{eq:prob_Si}) would have large fluctuation when there are few AGN
in the error volume.  

Therefore, we perform mock observations of GW events and AGNs to test whether a
non-zero $f_{\rm agn}$ could arise from statistical fluctuation.  First, we
simulate the case where there is no correlation between GW events and AGNs,
i.e., we inject $f^{\rm inj}_{\rm agn}=0$. More specifically, given each LVK
event, we look for the galaxies in the same redshift range from the SDSS galaxy
catalog \citep{2020ApJS..249....3A}. From these galaxies, we randomly pick one
as the mock host of the GW source. We then ``paste'' the skymap of the same GW
source to the location of the mock host, with a small displacement determined
by the probability $p_i(\mathbf{x}_j)$. After assigning mock hosts and
relocating the skymaps for all the LVK events, we can repeat the calculation
described in Section~\ref{sec:methodology} and derive a simulated $p(f_{\rm
agn})$ curve.

The results from $100$ such realizations are shown as the gray curves in the
left panels of Figure~\ref{fig:fagn_PDF_mock}. In either the lower-luminosity
case or the lower-Eddington-ratio one, we find that the simulated $p(f_{\rm
agn})$ curves mostly disagree with the real observational ones (cyan
curves). To further quantify the disagreement, we define a quantity 
which mimics the standard chi-square as follows
\begin{equation}  
\hat \chi^2 \equiv \int_0^1 \! \left( \frac{p_{\rm real}(f_{\rm agn}) - \bar p_{\rm mock}(f_{\rm agn})}{\sigma_p(f_{\rm agn})} \right)^{\!2}  \D f_{\rm agn},   \nonumber
\end{equation}
where $\bar p_{\rm mock}(f_{\rm agn})$ and $\sigma_p(f_{\rm agn})$ represent the median
and standard deviation of mock constraints, respectively.  Higher $\hat \chi^2$
values indicate larger inconsistency. 
When $f_{\rm agn}^{\rm inj} = 0$, we find $\hat \chi^2 \approx 0.2$
using the lower-$L_{\rm bol}$ sub-catalog of AGNs and $\hat \chi^2 \approx 0.4$
using lower-$\lambda_{\rm Edd}$ sub-catalog. 
For comparison, we inject $f_{\rm agn}^{\rm inj} = 0.4$ ($0.3$) and redo the
mock observation $100$ times using the lower-$L_{\rm bol}$ (lower-$\lambda_{\rm
Edd}$) AGN sub-catalog. The results are shown in the right panels of
Figure~\ref{fig:fagn_PDF_mock}. Now the $\hat \chi^2$ becomes $\approx0.1$
($0.08$).  The significant drop of the value of $\hat \chi^2$ is in favor of a
non-zero $f_{\rm agn}$ in real observation.

Alternatively, the relevance of statistical fluctuation can also be evaluated
through the cumulative probability distribution of $f_{\rm agn}$. 
For example, we can use
\begin{equation}  
\Delta P_{0.05} \!\equiv\! \int_0^{0.05} \! p(f_{\rm agn}) \, \D f_{\rm agn}   \nonumber
\end{equation}
to quantify the probability that $f_{\rm agn}$ is intrinsically small. We have
chosen a threshold of $f_{\rm agn}=0.05$ because observations suggest that
about $1\% - 10\%$ of galaxies are AGNs \citep{2007ApJ...654..731H,
2009MNRAS.399.1106M}, therefore $f_{\rm agn}=0.05$ is a typical value if LVK
events appear randomly in all types of galaxies.

Using real GW data and the 
lower-$L_{\rm bol}$ AGN sub-catalog (the cyan curves in the top panels of
Figure~\ref{fig:fagn_PDF_mock}), we find $\Delta P_{0.05}\simeq0.04$.
Therefore, the observational data do not favor a random coincidence between LVK events and lower-$L_{\rm bol}$ AGNs.
In comparison, if we inject
$f_{\rm agn}^{\rm inj} = 0$ and do mock observations, we find that 
$\Delta P_{0.05}$ will be higher than $0.04$ in 
about
$80\%$ of the realizations. 
The much higher $\Delta P_{0.05}$ values from the mock
simulations reject the $f_{\rm agn} = 0$ hypothesis at the $1.3\sigma$ credibility level.

Similarly, for the lower-$\lambda_{\rm Edd}$ AGN sub-catalog, we find
$\Delta P_{0.05}\simeq0.07$ using real GW data. It is smaller than the
$\Delta P_{0.05}$ values from about $90\%$ of the mock simulations. Therefore,
a random coincidence between LVK events and lower-$\lambda_{\rm Edd}$ AGNs is
rejected at a credibility of $1.6\sigma$.

\subsection{Origin of the signal}   \label{sec:result_poorLoc}

In our calculation, we find that large signal probabilities $\mathcal{S}_i$
often come from a few GW events with small localization error volumes ($\Delta
V_{\rm c}$). The number of AGNs within these error volumes is also small, which
is a potential source of statistical fluctuation (see
Appendix~\ref{sec:appendix_method}).  To understand the impact of such sources
on our estimation of $f_{\rm agn}$, we impose a threshold of $10^6~{\rm Mpc}^3$
and remove those GW events with smaller error volumes from our analysis. This
threshold corresponds to $O(1)$  AGN in the error volume, according to the
spatial density of SDSS AGNs in the lower-density sky regions.  

\begin{figure}[t]
 \centering
 \includegraphics[width=0.47\textwidth]{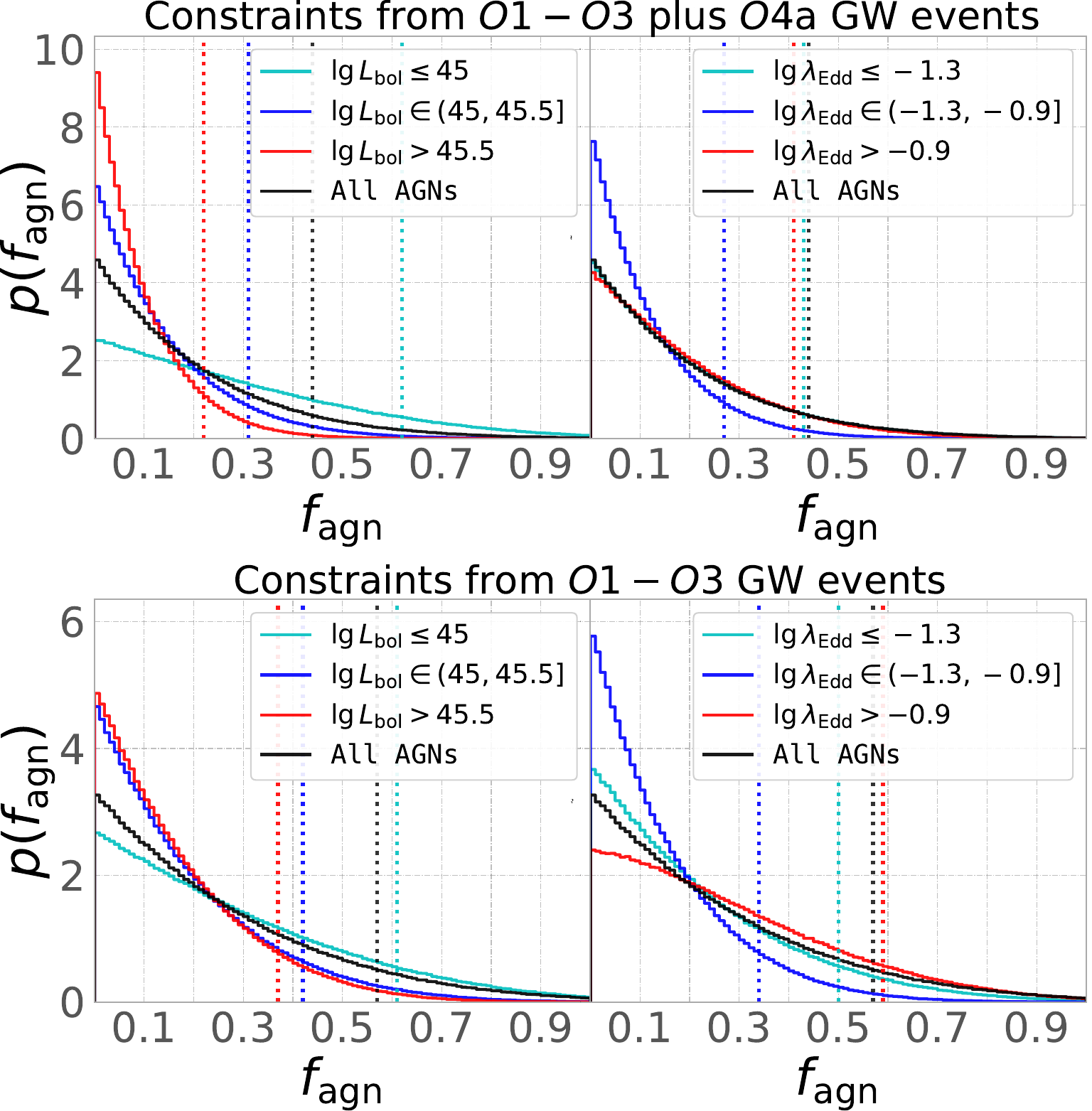}
 \caption{Probability distributions of $f_{\rm agn}$ derived from $O1\!-\!O4$a (top) 
 and $O1\!-\!O3$ (bottom) after excluding the four best localized LVK events. 
The line styles are the same
as in Figure~\ref{fig:fagn_PDF_allLocGW}.  The vertical dotted lines show the 
upper limits of $f_{\rm agn}$ according to the $90\%$ CL.
}
 \label{fig:fagn_PDF_poorLocGW}
\end{figure}

In our sample, only four $O4$ GW candidates (S230627c, S240413p, S250114ax, and
S250119cv) meet the above criterion.  While S230627c does not contain any
cataloged AGN in its error volume and hence does not contribute to the signal,
the latter three all contain AGNs in their error volumes, which have been shown
in Figure~\ref{fig:gw_skymaps}.  The inhomogeneity of the spatial probability
distribution as well as the few number of AGNs causes the value of
$\mathcal{S}_i$ to vary from $0.8$ for S250119cv to $3.9$ for S250114ax when
using the full AGN catalog.  The fluctuation is apparent. 

The top panels in Figure~\ref{fig:fagn_PDF_poorLocGW} show the probability
distribution of $f_{\rm agn}$ after we
remove the aforementioned four GW candidates. Now the most probable value of $f_{\rm agn}$
is zero in all the six sub-catalogs. This result suggests that the signal which we
found in the previous subsection regarding the correlation between GW events and AGNs
is predominated by the four most precisely localized events.

Although much weaker, the signal is not completely lost after we remove the
four best localized GW candidates. One can see this by comparing the current
results with those derived from only $O1\!-\!O3$ events, which are shown in the
bottom panels of Figure~\ref{fig:fagn_PDF_poorLocGW}. Since the $O1\!-\!O3$ GW
catalog is three times smaller than the full $O1\!-\!O4$ catalog, the
constraint on $f_{\rm agn}$ should be much weaker 
due to larger statistical fluctuation, and we should see 
a broader distribution of $f_{\rm agn}$ if there is intrinsically no
correlation between GW events and AGNs.  However, we find otherwise by
comparing the upper limits of $f_{\rm agn}$ (based on $90\%$ CL, see the
vertical dotted lines in Figure~\ref{fig:fagn_PDF_poorLocGW}). When we switch from the full GW catalog
to the  $O1\!-\!O3$-only catalog, the distribution of $f_{\rm agn}$ 
derived for the lower-$L_{\rm bol}$ sub-catalog becomes narrower, and the distribution for
the lower-$\lambda_{\rm Edd}$ catalog only slightly increases by a factor of $1.1$. While
in the other cases, the distribution broadens by $1.3$ to $1.7$ times.  
Such dependence on the number of GW events corroborates that 
LVK events are not randomly coinciding with
lower-luminosity or lower-Eddington-ratio AGNs. 

\section{Discussions}   \label{sec:discussion}

In this study, we quantified the fraction of the LVK \ac{GW} events which could
potentially originate from \acp{AGN}. 
We have improved the methodology presented in the previous works and 
divided AGNs into different groups according to their luminosities or
Eddington ratios. 
We find tentative evidence of GW events
coming from either the lower-luminosity 
($10^{44.5} \lesssim L_{\rm bol} \!<\! 10^{45}~\!\mathrm{erg~s}^{-1}$) 
or the lower-Eddington-ratio 
($0.01 \lesssim \lambda_{\rm Edd}\le0.05$) AGN populations. 
The fraction of these events is $f_{\rm agn} =
0.39^{+0.41}_{-0.32}$ in the former case and $0.29^{+0.40}_{-0.25}$ in the
latter one. Notice that we have been conservative in deriving these fractions
because we overestimated the completeness of AGN surveys 
(see the end of Section~\ref{sec:methodology}).  Our
simulated mock observations (Section~\ref{sec:result_preciLoc}) and the
analysis of sample size dependence (Section~\ref{sec:result_poorLoc})
further confirm that the signal is not due to statistical fluctuation
and remains consistent across different samples.

Our finding that a fraction of LVK GW events originate from the lower-luminosity
AGN population agrees with the earlier results \citep{veronesi23obs,
veronesi24obs}, which, using a similar methodology, demonstrated that GW events
cannot largely come from the most luminous AGNs. For example, using the
higher-luminosity AGN sub-catalog ($L_{\rm bol} > 10^{45.5}$ erg s$^{-1}$), we obtain
$f_{\rm agn}^{\rm best} = 0$ and an upper limit of $f_{\rm agn}^{\rm up} =
0.37$ ($90\%$ CL) for the $O1\!-\!O3$ events. These values become $f_{\rm
agn}^{\rm best} = 0$ and $f_{\rm agn}^{\rm up} = 0.16$ if we include $O4$
candidates into the analysis.  For comparison, the previous two studies
reported values of $f_{\rm agn}^{\rm up} = 0.40$ and $f_{\rm agn}^{\rm up} =
0.10$ \citep{veronesi23obs, veronesi24obs}, respectively. 
The differences can be attributed to our use of AGN catalogs, 
our selection criterion of GW events, and updated method of statistics. 
At redshift $z<1.5$, the SDSS AGN catalog used in our analysis and the Quaia catalog 
adopted by \cite{veronesi24obs} exhibit comparable survey-depth completeness. 
The key advantage of the SDSS catalog lies in its accurate spectroscopic-redshift measurements 
and homogeneous Eddington-ratio estimates \citep{2022ApJS..263...42W}. 
As for the Quaia catalog, although it uses relatively less reliable photo-redshifts, 
its main advantage is the larger sky area coverage \citep{2024ApJ...964...69S}. 
This expanded sky coverage of the Quaia catalog consequently allows 
the use of more GW candidates in their statistical analysis, e.g., 
$159$ GW candidates in \cite{veronesi24obs} versus $94$ GW candidates in our work.

Our findings are also consistent with multiple predictions made by the earlier
theoretical models about BBH formation.  For example, it has been proposed that
migration traps in AGN accretion disks could facilitate the formation of BBHs
\citep{2016ApJ...819L..17B, peng21}, but such traps may only exist in the AGNs with
a bolometric luminosity of $L_{\rm bol} < 10^{45}~\mathrm{erg~s}^{-1}$
\citep[assuming a viscous parameter of $\alpha_{\rm SS} =
0.01$][]{2024MNRAS.530.2114G} or in the narrower luminosity range of 
$10^{43.5}$--$10^{45.5}~\!\mathrm{erg~s}^{-1}$ \citep{2025ApJ...982L..13G}.
Moreover, there might be an anti-correlation between BBH merger rate
and the Eddington ratio if BBHs form in AGNs \citep{yang19}. This prediction is
also consistent with our findings. 

Besides the method used in this work, hierarchical Bayesian inference has been
commonly used to constrain the formation channels of the LVK GW events
\citep{2019MNRAS.486.1086M, 2019PASA...36...10T, gayathri21,
2022ApJ...941L..39W}.  This latter approach compares the observed distributions
of GW source parameters with the predictions from population models, and it is
less dependent on the completeness of AGN surveys but more on the details of
population models \citep[see][for different population
properties]{2007ApJ...661L.147B, mckernan12, 2018ApJ...866...66M, mckernan20,
2024arXiv241016515M, yang19mass-spin, tagawa20, 2022PhRvD.105f3006L,
2024PhRvL.133e1401L, 2024arXiv241110590C, 2024arXiv241107304A,
2025ApJ...979L..27L}.  In particular, \cite{2025ApJ...981..177L} recently
applied this method to $O1\!-\!O3$ events and found $f_{\rm agn} =
0.34_{-0.26}^{+0.38}$ at 90\% CL for hierarchical-merger GW candidates. 
Their result agrees remarkably well with ours, 
though our method can further reveal the AGN sub-populations 
(lower-$L_{\rm bol}$ or lower-$\lambda_{\rm Edd}$) which are responsible 
for producing the \ac{GW} sources.

Finally, we would like to emphasize that our constraints on $f_{\rm agn}$ will
improve significantly in the near future.  The current analysis uses $94$ GW
events from LVK's $O1-O4$ runs and an SDSS AGN catalog covering $\sim26\%$ of
the sky. Two key developments will soon enhance the precision of our
analysis: (i) the ongoing $O4$ run \citep{2020LRR....23....3A, LVK_plans} is
expected to yield $\sim$300 GW candidates, and (ii) new AGN catalogs covering
larger than $70\%$ of the sky are becoming available
\citep{2023OJAp....6E..49F, 2024ApJ...964...69S, 2024ApJS..271...54F}.  If
$f_{\rm agn} = 0.2$, assuming uncertainties scale as $1/\sqrt{{\rm SNR}}$
(``SNR'' stands for signal-to-noise ratio), we can expect a $\sim 50\%$
reduction in $f_{\rm agn}$ errors and a $\sim2$ time increase in the
significance of $f_{\rm agn} > 0$. These advances will help us better constrain
the formation channels of LVK GW events.

\section*{Acknowledgements}
The authors thank Yuming Fu, Linhua Jiang, 
Peng Peng, Qiaoya Wu, Yue Shen, Yi-Ming Hu, Han Wang, and Yu Liu 
for very helpful discussions. 
This work is supported by the National Key Research and Development Program of
China (Grant No. SQ2024YFC220046 and No. 2021YFC2203002). XC is supported by the National Natural Science Foundation of
China (Grant No. 12473037).
LGZ is funded by the China Postdoctoral Science Foundation (Grant No. 2023M740113). 

\software{\textsf{numpy} \citep{2011CSE....13b..22V}, 
\textsf{scipy} \citep{2020NatMe..17..261V}, 
\textsf{ligo.skymap} \citep{ligo_skymap}, 
\textsf{LALSuite} \citep{ligo_lalsuite}, 
\textsf{alphashape} \citep{alphashape_python}, 
\textsf{emcee} \citep{2013PASP..125..306F, 2019JOSS....4.1864F}, 
\textsf{matplotlib} \citep{2007CSE.....9...90H} and \textsf{seaborn} \citep{Waskom2021}. 
}

\appendix

\section{Notes on the data}   \label{sec:appendix_data}

This appendix provides detailed information about the GW events and AGN catalogs 
used in our analysis. Table~\ref{tab:gw_used} lists the 29 GW events from LVK's 
$O1 \!-\! O3$ runs and 65 $O4$ candidates, including their mean redshifts and 
comoving localization error volumes. 

\begin{figure}[tp]
 \centering
 \includegraphics[width=0.98\textwidth]{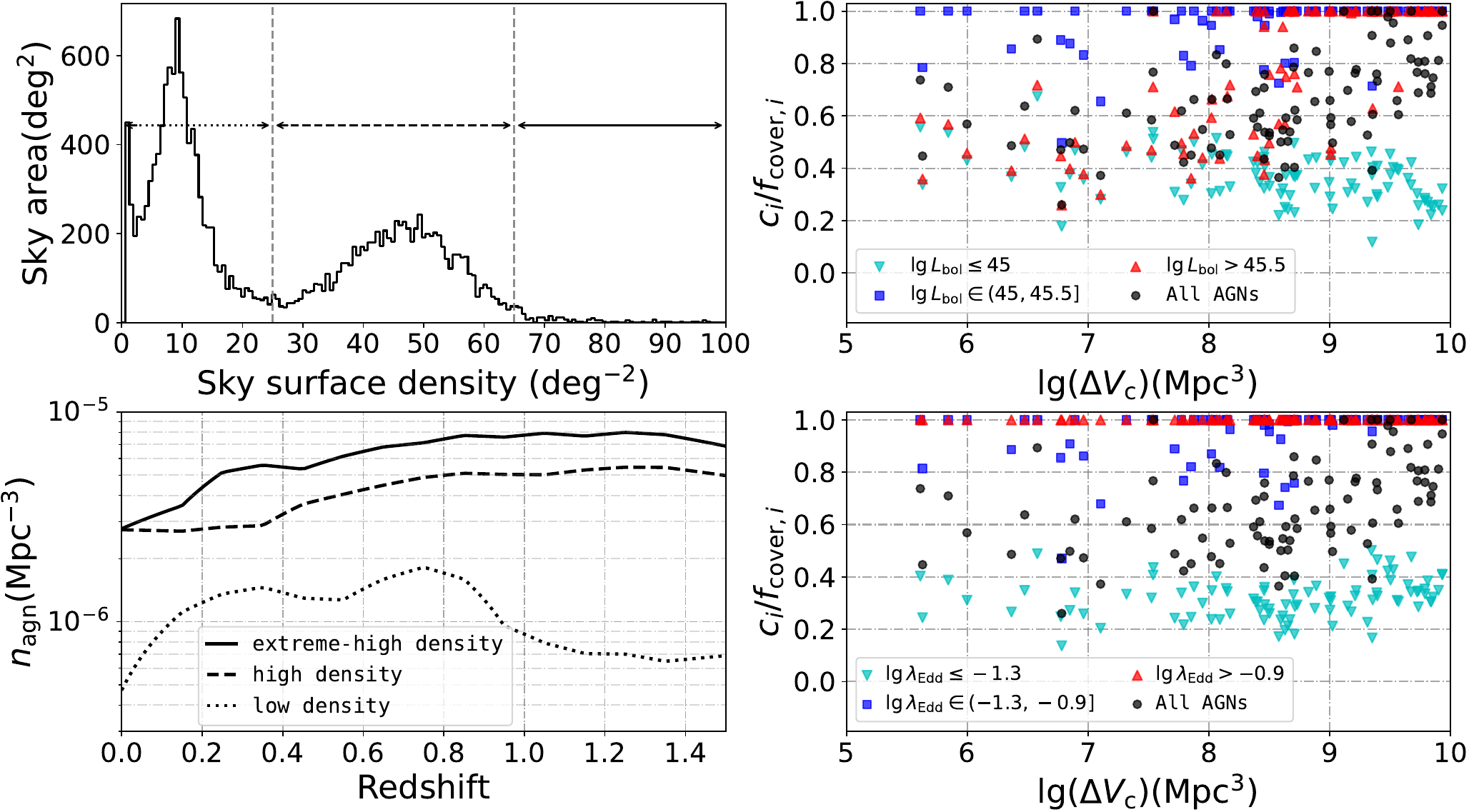}
 \caption{AGN density distributions and completeness analysis. 
 \textit{Top left}: On-sky surface density distribution of SDSS AGNs, showing cumulative area versus density. 
 The sky area element for statistics is $\Delta \Omega \approx 1.56~\mathrm{deg}^2$. 
 \textit{Bottom left}: Redshift-dependent spatial density of AGNs in  
 low (dotted line), high (dashed line), and extreme-high (solid line) density regions. 
 \textit{Right panels}: Survey-depth completeness ($c_i/f_{\mathrm{cover},i}$) versus 
 localization error volume ($\Delta V_{\mathrm{c}}$) for candidate AGN hosts of individual GW events, 
 color-coded by AGN sub-catalogs: black represents the full AGN catalog; 
 cyan, blue, and red, respectively, represent the lower, moderate, and 
 higher $L_{\mathrm{bol}}$ sub-catalogs (\textit{top}); 
 and corresponding $\lambda_{\mathrm{Edd}}$ divisions (\textit{bottom}). 
 }
 \label{fig:z_pdf}
\end{figure}

The SDSS AGN catalog exhibits significant spatial inhomogeneity. 
Figure~\ref{fig:z_pdf} (top left panel) shows the on-sky surface density distribution 
for SDSS AGNs with $z < 1.5$, which we categorize into three distinct regions: 
\begin{itemize}
    \item Extreme-high density region ($n_{\rm agn} > 65~\mathrm{deg}^{-2}$): located within $|\delta| \lesssim 1.25^\circ$ and $-40^\circ \lesssim \alpha \lesssim +45^\circ$;
    \item High density region ($25 < n_{\rm agn} \leq 65~\mathrm{deg}^{-2}$): covers declinations $+30^\circ \lesssim \delta \lesssim +60^\circ$;
    \item Low density region ($n_{\rm agn} \leq 25~\mathrm{deg}^{-2}$): comprises the remaining SDSS footprint.
\end{itemize}
The bottom left panel of Figure~\ref{fig:z_pdf} reveals substantial redshift-dependent 
density variations, particularly in low-density region where fluctuations approach 
one order of magnitude.

The survey-depth completeness of candidate AGN hosts exhibits significant variation across GW events, 
primarily due to differences in celestial coordinates and luminosity distances. 
The right panels in Figure \ref{fig:z_pdf} show the survey-depth completenesses 
($c_i / f_{\rm cover}$) for the GW events in Table \ref{tab:gw_used}, 
separating the full AGN catalog and 
the lower/moderate/higher $L_{\rm bol}$ and $\lambda_{\rm Edd}$ AGN sub-catalogs. 
Our completeness analysis is based on two observations results: (i) the bolometric luminosity function 
from \citet{2019MNRAS.488.1035K} with a cutoff $\lg L_{\rm bol} \geq 44.5$, 
and (ii) the Eddington ratio distribution function from \citet{2022ApJS..261....9A} 
with a cutoff $\lg \lambda_{\rm Edd} \geq -1.8$. 
The completeness for each GW event is defined as 
$c_i \equiv f_{{\rm cover},i} \times N_{\rm observed} / N_{\rm predicted}$ (also see Equation~(\ref{eq:compl})), 
where the predicted AGN number $N_{\rm predicted}$ is derived by integrating 
the respective distribution functions over the GW source's localization volume. 
We emphasize that the $L_{\rm bol}$ ($\lambda_{\rm Edd}$) cutoff strongly affects 
lower $L_{\rm bol}$ ($\lambda_{\rm Edd}$) AGN sub-catalog completeness 
estimates---lower cutoffs systematically reduce completeness values. 
Our chosen thresholds ($\lg L_{\rm bol}=44.5$ and $\lambda_{\rm Edd}= -1.8$) 
approximately correspond to the $5$th percentile of the AGN bolometric luminosity 
and Eddington ratio distributions in the SDSS AGN catalog used in this work. 
We note that this AGN catalog 
has not been corrected for Malmquist bias.

\begin{table*}[ht]
  \caption{Event IDs of the GW candidates used in this work, detected during the first three and partial fourth observing runs of LVK network, along with their mean redshifts and spatial localization error volumes $\Delta V_{\rm c}$ at the $90\%$ \ac{CL}.   }
  \vspace{-6pt}
  \renewcommand\arraystretch{1.3}
  \setlength\tabcolsep{-0.pt}
  \centering
  \begin{tabular}{|c|c|}
        \hline
        $O1 \!-\! O3$ GW candidate ($\Delta V_{\rm c}$ in ${\rm Mpc}^3$)  & $O4$a GW candidate ($\Delta V_{\rm c}$ in ${\rm Mpc}^3$)  \\
        \hline
        \!\!\!\!\!\!\!\!\!\!\!\!\!\!\!\!\!\thead[l]{
        GW170104$~\!        (\bar z=0.21, \Delta V_{\rm c}=1.42\times 10^{8})$, ~\\
        GW170608$~\!        (\bar z=0.07, \Delta V_{\rm c}= 2.98\times 10^{6})$, ~\\
        GW170818$~\!        (\bar z=0.20 , \Delta V_{\rm c}= 6.04\times 10^{6})$, ~\\
        GW190412\_053044$~\!(\bar z=0.15, \Delta V_{\rm c}= 9.16\times 10^{6})$, ~\\
        GW190425\_081805$~\!(\bar z=0.03, \Delta V_{\rm c}= 7.78\times 10^{6})$, ~\\
        GW190521\_030229$~\!(\bar z=0.48, \Delta V_{\rm c}= 3.02\times 10^{9})$, ~\\
        GW190630\_185205$~\!(\bar z=0.17, \Delta V_{\rm c}= 1.23\times 10^{8})$, ~\\
        GW190701\_203306$~\!(\bar z=0.36, \Delta V_{\rm c}= 3.46\times 10^{7})$, ~\\
        GW190706\_222641$~\!(\bar z=0.52, \Delta V_{\rm c}= 8.53\times 10^{9})$, ~\\
        GW190708\_232457$~\!(\bar z=0.18, \Delta V_{\rm c}= 1.02\times 10^{9})$, ~\\
        GW190803\_022701$~\!(\bar z=0.50 , \Delta V_{\rm c}= 2.21\times 10^{9})$, ~\\
        GW190915\_235702$~\!(\bar z=0.31, \Delta V_{\rm c}= 2.44\times 10^{8})$, ~\\
        GW190924\_021846$~\!(\bar z=0.11, \Delta V_{\rm c}= 1.27\times 10^{7})$, ~\\
        GW190925\_232845$~\!(\bar z=0.18, \Delta V_{\rm c}= 2.86\times 10^{8})$, ~\\
        GW190926\_050336$~\!(\bar z=0.47, \Delta V_{\rm c}= 7.94\times 10^{9})$, ~\\
        GW191103\_012549$~\!(\bar z=0.17, \Delta V_{\rm c}= 3.16\times 10^{8})$, ~\\
        GW191126\_115259$~\!(\bar z=0.28, \Delta V_{\rm c}= 6.66\times 10^{8})$, ~\\
        GW191127\_050227$~\!(\bar z=0.49, \Delta V_{\rm c}= 5.41\times 10^{9})$, ~\\
        GW191129\_134029$~\!(\bar z=0.15, \Delta V_{\rm c}= 5.92\times 10^{7})$, ~\\
        GW191204\_110529$~\!(\bar z=0.30 , \Delta V_{\rm c}= 4.44\times 10^{9})$, ~\\
        GW191219\_163120$~\!(\bar z=0.11, \Delta V_{\rm c}= 7.50\times 10^{7})$, ~\\
        GW200105\_162426$~\!(\bar z=0.06, \Delta V_{\rm c}= 3.35\times 10^{7})$, ~\\
        GW200115\_042309$~\!(\bar z=0.06, \Delta V_{\rm c}= 3.79\times 10^{6})$, ~\\
        GW200129\_065458$~\!(\bar z=0.17, \Delta V_{\rm c}= 7.06\times 10^{6})$, ~\\
        GW200202\_154313$~\!(\bar z=0.08, \Delta V_{\rm c}= 2.32\times 10^{6})$, ~\\
        GW200209\_085452$~\!(\bar z=0.51, \Delta V_{\rm c}= 2.47\times 10^{9})$, ~\\
        GW200210\_092254$~\!(\bar z=0.18, \Delta V_{\rm c}= 2.34\times 10^{8})$, ~\\
        GW200306\_093714$~\!(\bar z=0.35, \Delta V_{\rm c}= 4.69\times 10^{9})$, ~\\
        GW200311\_115853$~\!(\bar z=0.22, \Delta V_{\rm c}= 5.94\times 10^{6})$  }
        & \!\!\!\!\!\!\!\!\!\!\!\!\!\!\!\!\!\thead[l]{
        S230601bf$~\!(\bar  z=0.54, \Delta V_{\rm c}= 4.73\times 10^{9})$,~~\!~\!
        S230606d$~\! (\bar  z=0.31, \Delta V_{\rm c}= 7.73\times 10^{8})$, ~\\
        S230627c$~\! (\bar  z=0.06, \Delta V_{\rm c}= 4.26\times 10^{5})$,~~~\!~\! 
        S230628ax$~\! (\bar z=0.35, \Delta V_{\rm c}= 5.00\times 10^{8})$, ~~\\
        S230630bq$~\!(\bar  z=0.21, \Delta V_{\rm c}= 4.33\times 10^{8})$,~~\!\! 
        S230702an$~\!(\bar  z=0.41, \Delta V_{\rm c}= 3.16\times 10^{9})$, ~\!  \\
        S230704f$~\!(\bar   z=0.45, \Delta V_{\rm c}= 3.35\times 10^{9})$,~~~~\!\!
        S230706ah$~\! (\bar z=0.35, \Delta V_{\rm c}= 1.41\times 10^{9})$, ~~  \\
        S230707ai$~\!(\bar  z=0.56, \Delta V_{\rm c}= 6.87\times 10^{9})$,~~\!
        S230708bi$~\!(\bar  z=0.15, \Delta V_{\rm c}= 7.12\times 10^{7})$, ~  \\
        S230708t$~\!(\bar   z=0.50, \Delta V_{\rm c}= 5.28\times 10^{9})$,~~~\!  
        S230715bw$~\! (\bar z=0.13, \Delta V_{\rm c}= 1.02\times 10^{9})$, ~~  \\
        S230731an$~\!(\bar  z=0.20, \Delta V_{\rm c}= 1.06\times 10^{8})$, ~\!\!
        S230805x$~\!(\bar   z=0.56, \Delta V_{\rm c}= 6.14\times 10^{9})$, ~\\
        S230814ah$~\!(\bar  z=0.08, \Delta V_{\rm c}= 2.52\times 10^{8})$,~~\!\!
        S230904n$~\! (\bar  z=0.22, \Delta V_{\rm c}= 4.26\times 10^{8})$, ~~  \\
        S230919bj$~\!(\bar  z=0.30, \Delta V_{\rm c}= 4.68\times 10^{8})$,~~\!
        S231001aq$~\!(\bar  z=0.27, \Delta V_{\rm c}= 1.66\times 10^{9})$,~~\!~\! \\
        S231020ba$~\!(\bar  z=0.20, \Delta V_{\rm c}= 2.86\times 10^{8})$,~~\!\!
        S231028bg$~\!(\bar  z=0.57, \Delta V_{\rm c}= 2.45\times 10^{9})$, ~\\
        S231102w$~\! (\bar  z=0.59, \Delta V_{\rm c}= 7.16\times 10^{9})$, ~~\!\!
        S231104ac$~\!(\bar  z=0.25, \Delta V_{\rm c}= 2.89\times 10^{8})$,~~\!~\! \\
        S231108u$~\! (\bar  z=0.38, \Delta V_{\rm c}= 9.87\times 10^{8})$,~~~\!
        S231110g$~\!(\bar   z=0.33, \Delta V_{\rm c}= 7.52\times 10^{8})$, ~\\
        S231112ag$~\!(\bar  z=0.65, \Delta V_{\rm c}= 5.41\times 10^{9})$,~~\!
        S231113bw$~\!(\bar  z=0.25, \Delta V_{\rm c}= 5.14\times 10^{8})$, ~\\
        S231114n$~\! (\bar  z=0.33, \Delta V_{\rm c}= 1.05\times 10^{9})$,~~~
        S231123cg$~\!(\bar  z=0.21, \Delta V_{\rm c}= 5.36\times 10^{8})$,~~\!~\! \\
        S231206cc$~\!(\bar  z=0.23, \Delta V_{\rm c}= 1.05\times 10^{8})$,~~\!~\!
        S231213ap$~\!(\bar  z=0.45, \Delta V_{\rm c}= 3.18\times 10^{9})$, ~\\
        S231223j$~\!(\bar   z=0.31, \Delta V_{\rm c}= 2.05\times 10^{9})$, ~~~\!
        S231226av$~\!(\bar  z=0.20, \Delta V_{\rm c}= 3.45\times 10^{7})$,~~\! \\
        S231231ag$~\! (\bar z=0.20, \Delta V_{\rm c}= 3.70\times 10^{9})$,~~\!
        S240413p$~\! (\bar  z=0.10, \Delta V_{\rm c}= 9.93\times 10^{5})$, ~~   \\
        S240426s$~\!(\bar   z=0.09, \Delta V_{\rm c}= 8.84\times 10^{7})$, ~~~\!\!  
        S240501an$~\! (\bar z=0.68, \Delta V_{\rm c}= 6.46\times 10^{9})$,~~~~\!  \\
        S240505av$~\! (\bar z=0.59, \Delta V_{\rm c}= 6.12\times 10^{9})$, ~\!
        S240507p$~\!(\bar   z=0.24, \Delta V_{\rm c}= 5.06\times 10^{8})$,~~\!  \\
        S240512r$~\! (\bar  z=0.20, \Delta V_{\rm c}= 6.17\times 10^{7})$, ~~  
        S240515m$~\!(\bar   z=0.54, \Delta V_{\rm c}= 2.48\times 10^{9})$,~~\!  \\
        S240520cv$~\!(\bar  z=0.24, \Delta V_{\rm c}= 1.15\times 10^{8})$, ~\! 
        S240530a$~\!(\bar   z=0.22, \Delta V_{\rm c}= 3.17\times 10^{8})$,~~\!  \\
        S240601co$~\! (\bar z=0.25, \Delta V_{\rm c}= 3.79\times 10^{8})$, ~\! 
        S240615ea$~\!(\bar  z=0.53, \Delta V_{\rm c}= 2.24\times 10^{9})$,~~\!  \\
        S240621eb$~\!(\bar  z=0.64, \Delta V_{\rm c}= 6.89\times 10^{9})$, ~\!  
        S240622h$~\! (\bar  z=0.30, \Delta V_{\rm c}= 1.39\times 10^{8})$,~~~  \\
        S240627by$~\! (\bar z=0.26, \Delta V_{\rm c}= 4.09\times 10^{8})$,~~\!
        S240629by$~\!(\bar  z=0.23, \Delta V_{\rm c}= 5.21\times 10^{7})$, ~  \\
        S240807h$~\!(\bar   z=0.19, \Delta V_{\rm c}= 2.27\times 10^{9})$,~~~~\!\! 
        S240825ar$~\!(\bar  z=0.24, \Delta V_{\rm c}= 3.93\times 10^{8})$, ~\\
        S240915bd$~\!(\bar  z=0.13, \Delta V_{\rm c}= 2.90\times 10^{8})$,~~\!
        S240916ar$~\!(\bar  z=0.26, \Delta V_{\rm c}= 5.10\times 10^{8})$, ~\\
        S240920bz$~\!(\bar  z=0.30, \Delta V_{\rm c}= 2.63\times 10^{8})$,~~~\!\!
        S240920dw$~\!(\bar  z=0.18, \Delta V_{\rm c}= 2.08\times 10^{7})$, ~\\
        S240921cw$~\!(\bar  z=0.20, \Delta V_{\rm c}= 1.51\times 10^{9})$,~~\!
        S240923ct$~\! (\bar z=0.70, \Delta V_{\rm c}= 2.25\times 10^{9})$, ~~\\
        S241009an$~\!(\bar  z=0.22, \Delta V_{\rm c}= 1.49\times 10^{8})$, ~  
        S241102cy$~\!(\bar  z=0.46, \Delta V_{\rm c}= 1.62\times 10^{9})$,  \\
        S241113p$~\!(\bar   z=0.25, \Delta V_{\rm c}= 3.60\times 10^{9})$, ~~~\!
        S241130n$~\!(\bar   z=0.35, \Delta V_{\rm c}= 4.50\times 10^{8})$, ~~~\! \\
        S241230ev$~\!(\bar  z=0.76, \Delta V_{\rm c}= 8.53\times 10^{9})$, ~~\!\!
        S250109f$~\!(\bar   z=0.51, \Delta V_{\rm c}= 1.31\times 10^{9})$,  \\
        S250114ax$~\!(\bar  z=0.10, \Delta V_{\rm c}= 6.95\times 10^{5})$, ~
        S250118dp$~\!(\bar  z=0.36, \Delta V_{\rm c}= 1.00\times 10^{9})$,  \\
        S250119cv$~\!(\bar  z=0.10, \Delta V_{\rm c}= 4.08\times 10^{5})$ }  \\
		\hline
  \end{tabular}
  \label{tab:gw_used}
\end{table*}  

\section{Proof of the unbiasedness of our statistical framework}   \label{sec:appendix_method}

We demonstrate that our statistical framework (Section~\ref{sec:methodology}) provides an unbiased estimation of the fraction $f_{\rm agn}$ of GW events originating from AGNs. For the $i$-th GW event, define:
\begin{itemize}
    \item $\mathcal{B}_i$: background probability;
    \item $\Delta\mathcal{\bar S}_i \equiv \mathbb{E}[\mathcal{S}_i - \mathcal{B}_i]$: expected signal excess;
    \item $\Delta\mathcal{B}_i \equiv \mathcal{S}_i|_{\rm null} - \mathcal{B}_i$: stochastic fluctuation under null hypothesis.
\end{itemize}
The per-event likelihood combining AGN and non-AGN scenarios reads
\begin{align}  \label{eq:apd_Li}
\mathcal{L}_i &= \hat{f}_{\rm agn} \cdot (\mathcal{B}_i + \Delta\mathcal{\bar S}_i + \Delta\mathcal{B}_i ) + (1-\hat{f}_{\rm agn}) \cdot \mathcal{B}_i    \nonumber \\
&= \mathcal{B}_i + \hat{f}_{\rm agn} \cdot (\Delta\mathcal{\bar S}_i + \Delta\mathcal{B}_i ),
\end{align}
where $\hat{f}_{\rm agn} \equiv 0.9 \cdot c_i \cdot f_{\rm agn}$ 
incorporates the AGN catalog completeness $c_i$. 
For $N$ independent events, the joint likelihood becomes
\begin{align}
\mathcal{L} &= \prod_{i=1}^N \left[\mathcal{B}_i + \hat{f}_{\rm agn} \cdot \big( \Delta\mathcal{\bar S}_i  + \Delta\mathcal{B}_i \big)\right].  \label{eq:apd_Ltot}
\end{align}
Under the small-signal approximation ($\Delta\mathcal{B}_i,\Delta\mathcal{\bar S}_i \ll \mathcal{B}_i$), we expand to first order:
\begin{align}
\mathcal{L} &\approx \prod_{i=1}^N \mathcal{B}_i + \hat{f}_{\rm agn} \cdot 
\sum_{i=1}^N\left[ \big( \Delta\mathcal{\bar S}_i + \Delta\mathcal{B}_i \big) \cdot \prod_{j\neq i}\mathcal{B}_j\right].   \label{eq:apd_Ltot2}
\end{align}
Defining $\mathcal{B}_{\rm tot} \equiv \prod_{i=1}^N \mathcal{B}_i$, the above equation simplifies to
\begin{align}
\mathcal{L} &\approx \mathcal{B}_{\rm tot}  \cdot  \left[1 + \hat{f}_{\rm agn}  \cdot  \sum_{i=1}^N\frac{\Delta\mathcal{\bar S}_i + \Delta\mathcal{B}_i}{\mathcal{B}_i}\right].   \label{eq:apd_Ltot3}
\end{align}

Consider $n$ AGN-origin GW events among $N$ total events, the expectation value of the true likelihood is
\begin{align}
\bar{\mathcal{L}} &= \left[\prod_{i=1}^n \big(\mathcal{B}_i + \Delta\mathcal{\bar S}_i + \Delta\mathcal{B}_i  \big)\right] \cdot \left[\prod_{i=n+1}^N \mathcal{B}_i\right], \label{eq:apd_Ltot_n}
\end{align}
and a first-order expansion yields
\begin{align}
\bar{\mathcal{L}} &\approx \mathcal{B}_{\rm tot}  \cdot \left[1 + \sum_{i=1}^n \frac{\Delta\mathcal{\bar S}_i + \Delta\mathcal{B}_i}{\mathcal{B}_i}\right].  \label{eq:apd_Ltot_n2}
\end{align}
Equating with Eq.~\eqref{eq:apd_Ltot2} ($\mathcal{L} = \bar{\mathcal{L}}$) gives the estimator:
\begin{align}
\hat{f}_{\rm agn} &\approx \frac{\sum_{i=1}^n (\Delta\mathcal{\bar S}_i + \Delta\mathcal{B}_i)/\mathcal{B}_i}{\sum_{i=1}^N (\Delta\mathcal{\bar S}_i + \Delta\mathcal{B}_i)/\mathcal{B}_i}.   \label{eq:apd_hatf}
\end{align}
For large sample size ($n,N \gg 1$), the systematic signal predominates and we have
\begin{align}
\sum\Delta\mathcal{\bar S}_i/\mathcal{B}_i \gg 0, ~{\rm while}~ \sum\Delta\mathcal{B}_i/\mathcal{B}_i \to 0 .   \nonumber
\end{align}
Then we can get a simplified expression
\begin{align}
\hat{f}_{\rm agn} &\approx \frac{\sum_{i=1}^n \Delta\mathcal{\bar S}_i/\mathcal{B}_i}{\sum_{i=1}^N \Delta\mathcal{\bar S}_i/\mathcal{B}_i} 
+ \left[ 1 -  \frac{ \sum_{i=1}^n \big(\Delta \mathcal{\bar S}_i / \mathcal{B}_i \big)}{\sum_{i=1}^N \big(\Delta \mathcal{\bar S}_i / \mathcal{B}_i \big)}  \right] \!\cdot
\frac{ \sum_{i=1}^n \big(\Delta \mathcal{B}_{i} / \mathcal{B}_i \big)}{\sum_{i=1}^N \big(\Delta \mathcal{\bar S}_i / \mathcal{B}_i \big)}
+ \mathcal{O}\left[ \left(\frac{\sum \Delta\mathcal{B}_i/\mathcal{B}_i}{\sum \Delta\mathcal{\bar S}_i/\mathcal{B}_i}\right)^2 \right]. \label{eq:apd_hatf_approx}
\end{align}

Assuming uniform localization errors ($\Delta\mathcal{\bar S}_i/\mathcal{B}_i \approx {\rm const.}$), we obtain
\begin{align}
\mathbb{E}[\hat{f}_{\rm agn}] &= \frac{n}{N} \equiv f_{\rm agn},   \label{eq:apd_hatf_expect}
\end{align}
which confirms the estimator's unbiasedness. Moreover, the variance scales as
\begin{align}
{\rm Var}(\hat{f}_{\rm agn}) &\approx \left(1 - \frac{n}{N}\right)  \cdot  \frac{\sum_{i=1}^n \Delta\mathcal{B}_i/\mathcal{B}_i}{\sum_{i=1}^N \Delta\mathcal{\bar S}_i/\mathcal{B}_i}.   \label{eq:apd_hatf_std}
\end{align}
Based on the expression for variance, we find two scaling relations:
\begin{itemize}
    \item For fixed $f_{\rm agn}$: $\Delta\hat{f}_{\rm agn} \propto N^{-1/2}$ (standard statistical convergence);
    \item For fixed $N$: $\Delta\hat{f}_{\rm agn} \propto (1-f_{\rm agn})^{1/2}$.
\end{itemize}
The completeness factor $c_i$ requires independent calibration, but does not affect the estimator's statistical properties.

\end{CJK*}

\bibliography{references_sbhbAGN}
\end{document}